\documentclass[12pt]{article}
\usepackage[english]{babel}
\usepackage{amsmath,amsthm,amssymb,amsfonts}
\usepackage{color}
\newtheorem{theorem}{Theorem}

\begin{document}

\title{On the characterization\\ of constitutive equations for third grade \\viscous Korteweg fluids}
\author{M.~Gorgone and P.~Rogolino\\
\ \\
{\footnotesize Department of Mathematical and Computer Sciences,}\\
{\footnotesize Physical Sciences and Earth Sciences, University of Messina}\\
{\footnotesize Viale F. Stagno d'Alcontres 31, 98166 Messina, Italy}\\
{\footnotesize mgorgone@unime.it; progolino@unime.it}
}

\date{Published in \textit{Phys. Fluids} \textbf{33}, 043107 (2021).}

\maketitle

\begin{abstract}
We consider a model of a third grade viscous Korteweg--type fluid in three space dimensions, and apply the 
extended Liu procedure in order to explicitly solve the constraints imposed by the entropy principle on the 
non--local constitutive relations. We detail the algorithm we use, and are able to characterize the material 
functions involved in the constitutive equations. In a natural way, the application of the extended Liu 
procedure allows us to recover an extra term in the entropy flux, preserving all the features of third grade 
viscous Korteweg--type fluids. Moreover, a further constraint, in order to avoid that at equilibrium only 
very special phase boundaries are admissible, is investigated.
\end{abstract}


\section{Introduction}
\label{sec:intro}

In the framework of weakly non--local thermodynamics, Korteweg \cite{Korteweg} proposed a continuous model 
for fluids in which liquid and vapour phases may coexist. In particular, he introduced a constitutive 
relation for the stress tensor  $\mathbf{T}$  involving,  in its elastic part, the first and second order 
gradients of the mass density, in order to describe  the cohesive forces due to long--range interactions 
\cite{Heida-Malek}, 
\begin{equation}
\label{T}
\mathbf{T}=(-p+\alpha_1\Delta\rho+\alpha_2|\nabla\rho|^2)\mathbf{I}+
\alpha_3\nabla\rho\otimes\nabla\rho+\alpha_4\nabla\nabla\rho,
\end{equation}
where $p$ is the pressure, $\rho$ the mass density, $\mathbf{I}$ the identity matrix and $\alpha_i$ ($i=1,\ldots,4$) material coefficients depending on $\rho$; moreover, $\Delta$ is the laplacian operator, 
and the symbol $\otimes$ denotes tensorial product.
It is known that many substances, like geological materials, polymeric fluids, are among those 
that are able of 
flowing but which exhibit flow characteristics that cannot
be adequately described by the classical  viscous fluid model. In order to describe them, many 
models have been proposed. A  class of such material models consists of
what are now commonly referred to as fluids of differential type or, informally, as
Rivlin--Ericksen fluids \cite{Dunn_Raj1995}. In such kind of materials, only a very short  part of the history of the 
deformation gradient has an influence on the stress. More specifically, in an
incompressible fluid of differential type, apart from a constitutively indeterminate pressure, the
stress tensor is expressed as a function of the velocity gradient and its higher order time derivatives.
In fact, a continuum material is said to be of grade $N$ if the constitutive quantities are allowed to depend 
on all gradients of the deformation $\mathbf{F}$  equal to the integer $N$ \cite{True_Raja}. 
The constitutive relation for the Cauchy stress tensor of such a material is given by
\begin{equation}
\mathbf{T}=\mathbf{T}(\mathbf{F}, \dot{\mathbf {F}},\nabla\mathbf{F},\nabla^2\mathbf{F},\ldots,\nabla^N\mathbf{F)}.
\end{equation}

Korteweg--type fluids with the constitutive equation for Cauchy stress tensor given by relation (\ref{T}) belong to a subclass of materials of grade 3 in which the constitutive quantities  are allowed
to depend not only on the deformation $\mathbf{F}$ but also
on all gradients of the deformation less than or equal to the integer 3.
In recent years, such higher grade materials have been employed not only to model capillarity effects but 
also to analyze the structure of liquid--vapour phase transitions under both static 
\cite{aifa.ser1,aifa.ser2} and dynamic \cite{slem1,slem2} conditions. As observed by Dunn and Serrin 
\cite{DunnSerrin,Dunn}, these higher grade materials are, in general, incompatible with the restrictions 
placed by second law of thermodynamics \cite{Truesdell}. To overcome the complications due to the structure 
of Cauchy stress tensor of Korteweg fluids,  many efforts have been done  in order to ensure the 
compatibility with the second law of thermodynamics \cite{Heida-Malek,CST-JMP-2009,CST-JNET-2010,
COP-Elasticity-2011}.

As remarked in Refs.~\cite{DunnSerrin,Dunn}, in order to solve such an incompatibility, it has been  
introduced an additional rate of supply of mechanical energy, the interstitial working, suitable to model the 
long--range interactions between the molecules; in such a way, an energy extra--flux is included in the local 
balance of energy.  An alternative proposed method for ensuring the compatibility with second law of thermodynamics 
without  modifying the energy balance with the inclusion of extra--terms, requires to relax the classical 
form of entropy flux by including  an entropy extra--flux. {In fact, owing  to the non--locality 
expressed by the dependence on gradients  of mass density,  it is usual to formulate a general 
statement of second law in which the entropy flux is not the ratio of the heat flux and the absolute 
temperature \cite{muller}.  
These two possibilities are not equivalent and lead to different thermodynamical restrictions.
Therefore, the validity of one of these two choices should be derived 
on the basis of the kinetic theory or suitable experimental results.} 

In continuum thermodynamics, the entropy principle constitutes a basic tool in developing constitutive theories \cite{Truesdell}.
It requires the constitutive relations to be
such that all thermodynamical processes, \emph{i.e.}, all solutions of the field equations,
satisfy the entropy inequality.  
In the literature, there are two classical procedures for the exploitation of such a principle: the Coleman--Noll procedure \cite{colnol} and the Liu one \cite{liu}.

Recently, generalizations of these two classical techniques have
been introduced \cite{CST-JMP-2009,CST-JNET-2010,cim07,colnol esteso,COT-JMP-2011}.
These extended approaches, successfully used in many applications of physical interest (see, for instance, 
Refs.~\cite{CST-JMP-2009,CST-JNET-2010,COP-Elasticity-2011,COP-IJNLM-2013,COP-CMT-2015,OPR-2016,CGOP-2020,GOR-2020}), 
revealed essential in order to ensure the compatibility of non--local constitutive relations with second law 
of thermodynamics.
In particular, the extended Liu technique consists in subtracting  from the Clausius--Duhem inequality  a linear combination of the field equations and of the spatial 
gradients of the latter (in the following they are called \emph{extended equations}), up to the order of the 
gradients entering the state space. The coefficients of  this
linear combination are the Lagrange multipliers which may depend  on
the state space variables only.
Therefore, the number of independent constraints to be taken into
account is always equal to that of the unknown fields and of the gradients
entering the constitutive equations as independent state variables \cite{cim07}. 

{In several recent papers, special models of Korteweg--type fluids have been considered within the context of extended Liu technique for exploiting the the thermodynamical compatibility of the constitutive equations. In particular,  in Ref.~\cite{CST-JMP-2009}, it    
is proved that the  mathematical procedure above described allows to derive the consequences of the extended
entropy inequality for second grade  Korteweg--type viscous fluids.
Furthermore, in Ref.~\cite{GOR-2020}, it has been discussed the  extended Liu  procedure from an abstract mathematical point of view for a system of balance laws in one space dimension sufficiently  general to contain the equations governing the thermodynamical 
 processes occurring in a continuous medium; in particular, the results have been applied 
 to Korteweg fluids of grade 3.} Differently from the results obtained in Ref.~\cite{GOR-2020},
in the present paper, we study the thermodynamical  compatibility of the constitutive equations of a viscous third grade Korteweg fluid  in three space dimensions by means of the extended Liu procedure;
the approach we follow does not require neither the modification of the energy balance, that is taken in the classical form, nor an \emph{a priori} introduction of an entropy extra--flux. 
As a consequence of the algorithm 
whose details are given in the sequel, an entropy extra--flux naturally arises as a by--product of the 
mathematical exploitation of entropy principle with the extended Liu procedure. 
{
It is worth of observing that  in non--local theories, it is not uncommon to recognize the necessity of introducing a generalized expression of the  entropy flux due to physical considerations; typical examples fall in non--local fluid mechanical models \cite{morro2007}, phase--field models \cite{morro2006} for solidification and superconductivity \cite{Landau-Ginzburg}. }

The structure of the paper is as follows. In Section~\ref{sec:model}, we introduce the model for third grade  
 fluids and exploit the Clausius--Duhem inequality by means of the extended Liu procedure. Then, in 
 Section~\ref{sec:korteweg-fluid}, we restrict to the case of a viscous Korteweg--type  fluid, and provide 
 explicitly a solution for the constitutive equations. In  Section~\ref{sec:Serrin},  we consider a further 
 constraint to be imposed on the coefficients of the Cauchy stress tensor in order to avoid that the phase 
 boundaries at the equilibrium are constrained to very special configurations. Finally, in  
 Section~\ref{sec:conclusions}, we  discuss our results as well as  possible future developments.

\section{Thermodynamic restrictions for third grade  fluids}
\label{sec:model}

 In this Section, we consider a viscous third grade fluid, and investigate the restrictions placed by the entropy inequality in order to characterize the form of the non--local constitutive equations \cite{Truesdell_Noll}.
This class of fluids received a moderate attention in literature after the pioneering paper by Dunn and Serrin \cite{DunnSerrin}, in which the compatibility with the basic tenets of rational continuum thermodynamics \cite{Truesdell} has been extensively studied. In the last years, several authors proposed some generalizations \cite{CST-JMP-2009,CST-JNET-2010,COP-Elasticity-2011,cim07,colnol esteso,COT-JMP-2011}
by means of an extended Liu procedure in order to ensure the compatibility with second law of thermodynamics \cite{CST-JMP-2009,cim07,GOR-2020}, or using a different methodology  \cite{Heida-Malek}. 

Let $\mathcal{B}$ be a fluid occupying a compact and simply connected region $\mathcal{C}$ of a Euclidean point space $E^3$; at a continuum level, its evolution  is ruled by the field equations representing the local balances of mass, linear momentum and energy, respectively, 
\begin{equation}
\label{equations}
\begin{aligned}
&\mathcal{E}^{(1)}\equiv \frac{\partial \rho}{\partial t}+\nabla\cdot(\rho\mathbf{v})=0,\\
&\mathcal{E}^{(2)}\equiv \rho\left(\frac{\partial\mathbf{v}}{\partial t}+(\mathbf{v}\cdot\nabla)\mathbf{v}\right)-\nabla\cdot \mathbf{T}=\mathbf{0},\\
&\mathcal{E}^{(3)}\equiv\rho\left(\frac{\partial\varepsilon}{\partial t}+\mathbf{v}\cdot \nabla\varepsilon\right)-\mathbf{T}\cdot\nabla\mathbf{v}+\nabla\cdot\mathbf{q}=0,
\end{aligned}
\end{equation}
where $\rho(t,\mathbf{x})$ is the mass density, $\mathbf{v}(t,\mathbf{x})\equiv(v_1,v_2,v_3)$  the velocity,  $\varepsilon(t,\mathbf{x})$ the internal energy per unit mass, $\mathbf{T}$ the symmetric Cauchy stress tensor, and $\mathbf{q}$ the heat flux; here and in the sequel, $\mathbf{A}\cdot\mathbf{B}$ denotes the full contraction of vectors and tensors, \emph{i.e.}, $\mathbf{A}\cdot\mathbf{B}=\hbox{tr}(\mathbf{A}^T\cdot\mathbf{B})$, where $(\cdot)^T$ denotes transposition.
For the sake of simplicity, we have assumed the absence of body forces and heat sources.

Field equations~(\ref{equations}) must be closed by constitutive equations for the Cauchy stress tensor and heat 
flux in such a way the local entropy production
\begin{equation}
\label{entropy inequality}
\sigma_s=\rho\left(\frac{\partial s}{\partial t}+\mathbf{v}\cdot\nabla s\right)+\nabla\cdot \mathbf{J}
\end{equation}
be non--negative along any admissible thermodynamic process, $s$ being the specific entropy, and $\mathbf{J}$ 
the entropy flux;  $s$ and $\mathbf{J}$ are constitutive quantities  too. 

A constitutive theory requires the choice of the so called state variables; in view of the model we want to 
study, let us assume the state space spanned by the variables
\begin{equation}
\label{state space}
\mathcal{Z}=\{\rho, \varepsilon, \nabla \rho, \mathbf{L} , \nabla\varepsilon, \nabla\nabla\rho \},
\end{equation}
where $\mathbf{L}=\hbox{sym}(\nabla\mathbf{v})=\dot{\mathbf F}\mathbf F^{-1}$; thus,
we put ourselves in the framework of a second order non--local constitutive theory.
{In the present paper, we analyze the class of Korteweg--type materials 
described by the  set of constitutive equations,
\begin{equation}
\mathcal{F}=\mathcal{F}^*(\rho, \varepsilon, \nabla \rho, \mathbf{L} , \nabla\varepsilon, \nabla\nabla\rho ),
\end{equation}
where $\mathcal{F}$ is an element of the set $\{\mathbf{T}, \mathbf{q},\mathbf{J}, s\}$. 
 It is worth of being remarked that the  principle of material frame indifference \cite{truesdellfirst}  
 implies that the elements of $\mathcal{F}$ are independent of the velocity field $\mathbf{v}$, and may depend on the velocity gradient only through its symmetric part; moreover, some restrictions on the  representation of isotropic constitutive quantities must be considered \cite{Smith,Wang1,Wang2}.}
 
Since we are considering a non--local theory, the exploitation of second law of 
thermodynamics must be done by means of the Liu's extended procedure \cite{cim07,colnol 
esteso,COT-JMP-2011}, whose general scheme,  in the case of $r$--th order non--local 
constitutive equations for one--dimensional continuous media, has been analyzed in 
Ref.~\cite{GOR-2020}.
Thus, it is necessary to take into account the constraints imposed on the thermodynamic 
processes by the balance equations together with the first and second order gradients 
of mass continuity equation, and the first order gradients of linear momentum and energy balance 
equations.
{
The necessity of imposing as additional constraints in the entropy inequality the gradients
of the balance equations when dealing with non--local constitutive
equations has stringent mathematical motivations. In fact,
the thermodynamic processes are solutions of the balance equations, and, if these solutions are smooth 
enough, are trivially solutions of their differential consequences \cite{RogCim}. Since the entropy 
inequality has to be satisfied also in arbitrary smooth processes, from a mathematical point of view, we have to 
use the differential consequences of the equations governing those processes as constraints for such an 
inequality. On the contrary, if we restrict ourselves to consider as constraints only the balance equations, 
we are led straightforwardly to a specific entropy and Lagrange multipliers which are independent of the 
gradients of the field variables entering the state space. As a direct consequence, a Cauchy stress tensor 
depending on the gradients of mass density would become incompatible with second law of thermodynamics.}
Therefore, in order to perform the exploitation of entropy inequality, 
let us introduce the Lagrange multipliers $\lambda^{(1)}$, $\boldsymbol\lambda^{(2)}$ 
(with components 
$\lambda^{(2)}_i$, $i=1,2,3$), $\lambda^{(3)}$, $\boldsymbol\Lambda^{(1)}$ (with 
components $\Lambda^{(1)}_i$, $i=1,2,3$), 
$\boldsymbol\Lambda^{(2)}$ (with components $\Lambda^{(2)}_{ij}$, $i,j=1,2,3$), $\boldsymbol\Lambda^{(3)}$  
(with components $\Lambda^{(3)}_{i}$, $i=1,2,3$) and $\boldsymbol\Theta^{(1)}$  (with components $
\Theta^{(1)}_{ij}$, $i,j=1,2,3$), depending on the state variables.
Thus,  the entropy inequality writes
\begin{equation}
\label{entropyconstrained}
\begin{aligned}
&\rho\left(\frac{\partial s}{\partial t}+\mathbf{v}\cdot\nabla s\right)+\nabla\cdot \mathbf{J}\\
&\qquad -\lambda^{(1)} \left(\frac{\partial \rho}{\partial t}+\nabla\cdot(\rho\mathbf{v}) \right)\\
&\qquad -\boldsymbol\lambda^{(2)}\cdot \left(\rho\left(\frac{\partial\mathbf{v}}{\partial t}+(\mathbf{v}\cdot\nabla)\mathbf{v}\right)-\nabla\cdot \mathbf{T} \right)\\
&\qquad -\lambda^{(3)} \left(\rho\left(\frac{\partial\varepsilon}{\partial t}+\mathbf{v}\cdot \nabla\varepsilon\right)-\mathbf{T}\cdot\nabla\mathbf{v}+\nabla\cdot\mathbf{q} \right)\\
&\qquad-\boldsymbol\Lambda^{(1)}\cdot \nabla\left(\frac{\partial \rho}{\partial t}+\nabla\cdot(\rho\mathbf{v}) \right)\\
&\qquad -\boldsymbol\Lambda^{(2)}\cdot \nabla\left(\rho\left(\frac{\partial\mathbf{v}}{\partial t}+(\mathbf{v}\cdot\nabla)\mathbf{v}\right)-\nabla\cdot \mathbf{T} \right)\\
&\qquad -\boldsymbol\Lambda^{(3)}\cdot \nabla\left(\rho\left(\frac{\partial\varepsilon}{\partial t}+\mathbf{v}\cdot \nabla\varepsilon\right)-\mathbf{T}\cdot\nabla\mathbf{v}+\nabla\cdot\mathbf{q} \right)\\
&\qquad -\boldsymbol\Theta^{(1)} \cdot \nabla\left(\nabla\left(\frac{\partial \rho}{\partial t}+\nabla\cdot(\rho\mathbf{v}) \right)\right)\ge 0.
\end{aligned}
\end{equation}

In condition (\ref{entropyconstrained}), we have to expand derivatives with the chain rule; the long computations, though straightforward, are done using some routines written in the Computer Algebra System Reduce \cite{Reduce}, allowing us to extract the coefficients of a multivariate polynomial in some derivatives of the field variables, and then solve, with the help of Crack package \cite{Wolf}, the set of differential and algebraic conditions for the unknown constitutive functions.  In components, the inequality (\ref{entropyconstrained}) can be written under the form
\begin{equation}
\label{entropyconstrained2}
\begin{split}
&\left(\rho\frac{\partial s}{\partial\rho}-\lambda^{(1)}\right)\rho_{,t}-\left(\rho_{,k}\Lambda_{ik}^{(2)}+\rho\lambda^{(2)}_i\right)v_{i,t}\\
&\quad+\left(\rho\frac{\partial s}{\partial\varepsilon}-\rho_{,k}\Lambda_k^{(3)}-\rho\lambda^{(3)}\right)\varepsilon_{,t}\\
&\quad+\left(\rho\frac{\partial s}{\partial\rho_{,k}}-\Lambda_k^{(1)}\right)\rho_{,kt}+\left(\rho\frac{\partial s}{\partial v_{i,k}}-\rho\Lambda_{ik}^{(2)}\right)v_{i,kt}\\
&\quad+\left(\rho\frac{\partial s}{\partial\varepsilon_{,k}}-\rho\Lambda_k^{(3)}\right)\varepsilon_{,kt}+\left(\rho\frac{\partial s}{\partial\rho_{,ik}}-\Theta_{ik}^{(1)}\right)\rho_{,ikt}\\
&\quad+\left(\Lambda_{ik}^{(2)}\frac{\partial T_{ij}}{\partial v_{n,m}}-\Lambda_k^{(3)}\frac{\partial q_j}{\partial v_{n,m}}-\rho\Theta_{km}^{(1)}\delta_{jn}\right)v_{n,jkm}\\
&\quad+\left(\Lambda_{ik}^{(2)}\frac{\partial T_{ij}}{\partial\varepsilon_{,m}}-\Lambda_k^{(3)}\frac{\partial q_j}{\partial\varepsilon_{,m}}\right)\varepsilon_{,jkm}\\
&\quad+\left(\Lambda_{ik}^{(2)}\frac{\partial T_{ij}}{\partial\rho_{,mn}}-\Lambda_k^{(3)}\frac{\partial q_j}{\partial\rho_{,mn}}\right)\rho_{,jkmn}\\
&\quad+f(\rho,v_i,\varepsilon,\rho_{,i},v_{i,j},\varepsilon_{,i},\rho_{,ij},v_{i,jk},\varepsilon_{,ij},\rho_{,ijk})\geq0,
\end{split}
\end{equation}
where $\delta_{ij}$ is the Kronecker delta, the Einstein convention on sums over repeated indices has been used, and the subscripts $(\cdot)_{,t}$ and $(\cdot)_{,j}$ stand for partial derivatives with respect to the time $t$  and the spatial coordinate $x_j$; moreover, because of its length, we omit to write the expression of the function $f$.

In relation (\ref{entropyconstrained2}), we can distinguish the \emph{highest derivatives} and the \emph{higher derivatives}  \cite{CST-JMP-2009};
highest derivatives are both the time derivatives of the field and 
state space variables, which cannot be expressed through the governing equations as functions of the 
thermodynamic variables, and the spatial derivatives with highest order, whereas higher derivatives are the spatial derivatives whose order is not maximal but higher than that of the gradients entering the state space. 

In the present case,  the highest derivatives are the elements of the set
\[
\{\rho_{,t},v_{i,t},\varepsilon_{,t},\rho_{,it},v_{i,jt},\varepsilon_{,it},\rho_{,ijt},
,v_{i,jk\ell},\varepsilon_{,ijk},\rho_{,ijk\ell}\},
\]
whereas the higher derivatives are the elements of the set
\[
\left\{v_{i,jk},\varepsilon_{,ij},\rho_{,ijk}\right\}.
\]
Inequality (\ref{entropyconstrained2}) results linear in the highest derivatives
with coefficients depending at most on the field and state space variables. From a mathematical point of view, since these quantities are independent of the elements of the state space and can assume
arbitrary values \cite{colnol,liu}, their coefficients must vanish, otherwise the above inequality could be easily violated. This leads to obtain the following expressions for the Lagrange multipliers, namely
\begin{equation}
\label{multipliers}
\begin{aligned}
&\lambda^{(1)}=\rho\frac{\partial s}{\partial\rho},\qquad \lambda^{(2)}_i=-\frac{\rho_{,k}}{\rho}\frac{\partial s}{\partial v_{i,k}},\qquad \lambda^{(3)}=\frac{\partial s}{\partial\varepsilon}-\frac{\rho_{,k}}{\rho}\frac{\partial s}{\partial\varepsilon_{,k}},\\
&\Lambda_{k}^{(1)}=\rho\frac{\partial s}{\partial\rho_{,k}},\qquad \Lambda_{ik}^{(2)}=\frac{\partial s}{\partial v_{i,k}},\qquad \Lambda_{k}^{(3)}=\frac{\partial s}{\partial\varepsilon_{,k}},\\
&\Theta_{ik}^{(1)}=\rho\frac{\partial s}{\partial\rho_{,ik}};
\end{aligned}
\end{equation}
using relations (\ref{multipliers}), the remaining coefficients of the highest derivatives provide the following thermodynamic restrictions involving the entropy, the Cauchy stress tensor and the heat flux:
\begin{equation}
\label{restrhighest}
\begin{aligned}
&\left\langle\frac{\partial s}{\partial v_{i,k}}\frac{\partial T_{ij}}{\partial v_{n,m}}\right\rangle_{(jkm)}=\left\langle\frac{\partial s}{\partial\varepsilon_{,k}}\frac{\partial q_j}{\partial v_{n,m}}+\rho^2\frac{\partial s}{\partial\rho_{,km}}\delta_{jn}\right\rangle_{(jkm)},\\
&\left\langle\frac{\partial s}{\partial v_{i,k}}\frac{\partial T_{ij}}{\partial\varepsilon_{,m}}\right\rangle_{(jkm)}=\left\langle\frac{\partial s}{\partial\varepsilon_{,k}}\frac{\partial q_j}{\partial\varepsilon_{,m}}\right\rangle_{(jkm)},\\
&\left\langle\frac{\partial s}{\partial v_{i,k}}\frac{\partial T_{ij}}{\partial\rho_{,mn}}\right\rangle_{(jkmn)}=\left\langle\frac{\partial s}{\partial\varepsilon_{,k}}\frac{\partial q_j}{\partial\rho_{,mn}}\right\rangle_{(jkmn)},
\end{aligned}
\end{equation}
where  the symbol $\left\langle\mathcal F\right\rangle_{(i_1\ldots i_r)}$ denotes the symmetric part of the tensor function $\mathcal F$  with respect to the indices $i_1\ldots i_r$.
It can be easily ascertained that the entropy inequality, that now reduces to
\begin{equation}
f(\rho,v_i,\varepsilon,\rho_{,i},v_{i,j},\varepsilon_{,i},\rho_{,ij},v_{i,jk},\varepsilon_{,ij},\rho_{,ijk})\geq0,
\end{equation}
is quadratic in the higher derivatives, and so it can be written in the form 
\begin{equation}\label{residualquad}
\mathbf{Y}^T \mathbf{A}\mathbf{Y} + \mathbf{B}\mathbf{Y} + C\geq0,
\end{equation}
where $\mathbf{Y} = (v_{i,jk},\varepsilon_{,ij},\rho_{,ijk})^T$ denotes the vector of higher derivatives, 
$\mathbf{A}$ is a symmetric matrix, $\mathbf{B}$ is a vector, and $C$ a scalar; $\mathbf{A}$, $\mathbf{B}$ and $C$ depend at most on the field and state space variables. 
Let us observe that nothing prevents to have a thermodynamic process where $C=0$. Moreover, since we used in the entropy inequality all the constraints imposed by the field equations together with their spatial derivatives, the higher derivatives may assume arbitrary values.
In order to analyze the relation (\ref{residualquad}), let us recall the Theorem proved in Ref. \cite{COT-JMP-2011}.
\begin{theorem}
The inequality
\begin{equation}\label{residua quadraticabis}
\mathbf{Y}^T \mathbf{A}\mathbf{Y} + \mathbf{B}\mathbf{Y} + C\geq0,
\end{equation}
where $\mathbf{A}$ is a $n\times n$ matrix, $\mathbf{B}\in\mathbb{R}^n$ and
$C\in\mathbb{R}$,
holds for arbitrary $\mathbf{Y}\in\mathbb{R}^n$, where $\mathbf{Y}$ can be chosen in such a way $C=0$, if and only if $\mathbf{A}$ is positive semidefinite, $\mathbf{B}\equiv\mathbf{0}$ and $C\geq0$.
\end{theorem}
 The requirement $\mathbf{B}\equiv 0$ in relation (\ref{residualquad}), \emph{i.e}, the coefficients of linear terms in the higher derivatives must be vanishing, yields the following set of additional thermodynamic restrictions:
\begin{equation}
\label{restrhigherv}
\begin{aligned}
\left\langle\frac{\partial J_j}{\partial v_{i,k}}\right\rangle
&=\left\langle-\rho v_j\frac{\partial s}{\partial v_{i,k}}
-\lambda^{(2)}_\ell\frac{\partial T_{\ell j}}{\partial v_{i,k}}+\lambda^{(3)}\frac{\partial q_j}{\partial v_{i,k}}+\rho\Lambda_{k}^{(1)}\delta_{ij}\right.\\
&+\rho_{,i}\Theta_{jk}^{(1)}+2\rho_{,\ell}\Theta_{\ell k}^{(1)}\delta_{ij}\\
&-\Lambda_{mk}^{(2)}\left(\frac{\partial^2 T_{\ell m}}{\partial\rho\partial v_{i,j}}\rho_{,\ell}
+\frac{\partial^2 T_{\ell m}}{\partial\varepsilon\partial v_{i,j}}\varepsilon_{,\ell}
+\frac{\partial^2 T_{\ell m}}{\partial\rho_{,n}\partial v_{i,j}}\rho_{,\ell n}-\rho\delta_{im}v_j
\right)\\
&-\Lambda_{\ell m}^{(2)}\left(\frac{\partial^2 T_{\ell j}}{\partial\rho\partial v_{i,k}}\rho_{,m}
+\frac{\partial^2 T_{\ell j}}{\partial\varepsilon\partial v_{i,k}}\varepsilon_{,m}
+\frac{\partial^2 T_{\ell j}}{\partial\rho_{,n}\partial v_{i,k}}\rho_{,m n}
\right)\\
&+\Lambda_{k}^{(3)}\left(\frac{\partial^2 q_\ell}{\partial\rho\partial v_{i,j}}\rho_{,\ell}
+\frac{\partial^2 q_\ell}{\partial\varepsilon\partial v_{i,j}}\varepsilon_{,\ell}
+\frac{\partial^2 q_\ell}{\partial\rho_{,m}\partial v_{i,j}}\rho_{,\ell m}-\frac{\partial T_{\ell m}}{\partial v_{i,j}}v_{\ell,m}-T_{ij}
\right)\\
&+\left.\Lambda_{\ell}^{(3)}\left(\frac{\partial^2 q_k}{\partial\rho\partial v_{i,j}}\rho_{,\ell}
+\frac{\partial^2 q_k}{\partial\varepsilon\partial v_{i,j}}\varepsilon_{,\ell}
+\frac{\partial^2 q_k}{\partial\rho_{,m}\partial v_{i,j}}\rho_{,\ell m}
\right)\right\rangle,
\end{aligned}
\end{equation}
\begin{equation}
\label{restrhigherepsilon}
\begin{aligned}
\left\langle\frac{\partial J_j}{\partial\varepsilon_{,k}}\right\rangle
&=\left\langle-\rho v_j\frac{\partial s}{\partial \varepsilon_{,k}}
-\lambda^{(2)}_i\frac{\partial T_{ij}}{\partial \varepsilon_{,k}}+\lambda^{(3)}\frac{\partial q_j}{\partial\varepsilon_{,k}}\right.\\
&-\Lambda_{ik}^{(2)}\left(\frac{\partial^2 T_{i\ell}}{\partial\rho\partial \varepsilon_{,j}}\rho_{,\ell}
+\frac{\partial^2 T_{i\ell}}{\partial\varepsilon\partial\varepsilon_{,j}}\varepsilon_{,\ell}
+\frac{\partial^2 T_{i\ell}}{\partial\rho_{,m}\partial\varepsilon_{,j}}\rho_{,\ell m}+\frac{\partial T_{ij}}{\partial \varepsilon}
\right)\\
&-\Lambda_{i\ell}^{(2)}\left(\frac{\partial^2 T_{i k}}{\partial\rho\partial \varepsilon_{,j}}\rho_{,\ell}
+\frac{\partial^2 T_{i k}}{\partial\varepsilon\partial\varepsilon_{,j}}\varepsilon_{,\ell}
+\frac{\partial^2 T_{i k}}{\partial\rho_{,m}\partial\varepsilon_{,j}}\rho_{,\ell m}
\right)\\
&+\Lambda_{k}^{(3)}\left(\frac{\partial^2 q_\ell}{\partial\rho\partial \varepsilon_{,j}}\rho_{,\ell}
+\frac{\partial^2 q_\ell}{\partial\varepsilon\partial\varepsilon_{,j}}\varepsilon_{,\ell}
+\frac{\partial^2 q_\ell}{\partial\rho_{,m}\partial\varepsilon_{,j}}\rho_{,\ell m}+\frac{\partial q_j}{\partial \varepsilon}-\frac{\partial T_{\ell m}}{\partial\varepsilon_{,j}}v_{\ell,m}+\rho v_j
\right)\\
&+\left.\Lambda_{\ell}^{(3)}\left(\frac{\partial^2 q_k}{\partial\rho\partial \varepsilon_{,j}}\rho_{,\ell}
+\frac{\partial^2 q_k}{\partial\varepsilon\partial\varepsilon_{,j}}\varepsilon_{,\ell}
+\frac{\partial^2 q_k}{\partial\rho_{,m}\partial\varepsilon_{,j}}\rho_{,\ell m}
\right)\right\rangle,
\end{aligned}
\end{equation}
\begin{equation}
\label{restrhigherrho}
\begin{aligned}
\left\langle\frac{\partial J_j}{\partial\rho_{,ik}}\right\rangle
&=\left\langle-\rho v_j\frac{\partial s}{\partial \rho_{,ik}}
-\lambda^{(2)}_\ell\frac{\partial T_{\ell j}}{\partial \rho_{,ik}}+\lambda^{(3)}\frac{\partial q_j}{\partial\rho_{,ik}}+\Theta_{ik}^{(1)}v_j\right.\\
&-\Lambda_{mk}^{(2)}\left(\frac{\partial^2 T_{\ell m}}{\partial\rho\partial \rho_{,ij}}\rho_{,\ell}
+\frac{\partial^2 T_{\ell m}}{\partial\varepsilon\partial\rho_{,ij}}\varepsilon_{,\ell}
+\frac{\partial^2 T_{\ell m}}{\partial\rho_{,n}\partial\rho_{,ij}}\rho_{,\ell n}+\frac{\partial T_{jm}}{\partial \rho_{,i}}
\right)\\
&-\Lambda_{\ell m}^{(2)}\left(\frac{\partial^2 T_{\ell k}}{\partial\rho\partial \rho_{,ij}}\rho_{,m}
+\frac{\partial^2 T_{\ell k}}{\partial\varepsilon\partial\rho_{,ij}}\varepsilon_{,m}
+\frac{\partial^2 T_{\ell k}}{\partial\rho_{,n}\partial\rho_{,ij}}\rho_{,m n}
\right)\\
&+\Lambda_{k}^{(3)}\left(\frac{\partial^2 q_\ell}{\partial\rho\partial \rho_{,ij}}\rho_{,\ell}
+\frac{\partial^2 q_\ell}{\partial\varepsilon\partial\rho_{,ij}}\varepsilon_{,\ell}
+\frac{\partial^2 q_\ell}{\partial\rho_{,m}\partial\rho_{,ij}}\rho_{,\ell m}+\frac{\partial q_j}{\partial \rho_{,i}}-\frac{\partial T_{\ell m}}{\partial\rho_{,ij}}v_{\ell,m}
\right)\\
&+\left.\Lambda_{\ell}^{(3)}\left(\frac{\partial^2 q_k}{\partial\rho\partial \rho_{,ij}}\rho_{,\ell}
+\frac{\partial^2 q_k}{\partial\varepsilon\partial\rho_{,ij}}\varepsilon_{,\ell}
+\frac{\partial^2 q_k}{\partial\rho_{,m}\partial\rho_{,ij}}\rho_{,\ell m}
\right)\right\rangle,
\end{aligned}
\end{equation}
where  the symbol $\left\langle\mathcal F_{(i_1\ldots i_r)}\right\rangle$ denotes the symmetric part respect to all  the indices  of $\mathcal F$.

Conditions (\ref{restrhighest}), and (\ref{restrhigherv})--(\ref{restrhigherrho}) provide restrictions on the constitutive functions. 
In particular, looking at Eqs. (\ref{multipliers}) and (\ref{restrhighest}), it is recognized that the Lagrange multipliers and, hence, the specific entropy can depend on the gradients of all the unknown variables, and that the same is true for the Cauchy stress tensor.
Furthermore, from Eqs. (\ref{restrhigherv})--(\ref{restrhigherrho}) it follows that, for the class of fluids at hand, the entropy flux is no longer given by the constitutive equation postulated in rational thermodynamics, namely
$\mathbf{J}=\displaystyle\frac{\mathbf{q}}{\theta}$, and an entropy extra--flux can be obtained.
Although the relations  (\ref{restrhighest}) and (\ref{restrhigherv})--(\ref{restrhigherrho}) place severe restrictions on the form
of the constitutive functions, they are still too much general for practical applications;
therefore, a further simplification is necessary according to specific models in order to solve all these conditions and to provide an explicit solution satisfying the remaining restrictions expressed as inequalities. 

\section{The case of Korteweg--type fluids}
\label{sec:korteweg-fluid}
In this Section, we assume a relation for Cauchy stress tensor generalizing the one proposed by Korteweg \cite{Korteweg}, and take the heat flux linear in the gradients of mass density and internal energy. In such a way, we are able to develop a complete analysis of the compatibility of these constitutive equations with the entropy principle.
Therefore, let us assume:
\begin{eqnarray}
\mathbf{T}&=&\left(-p+\alpha_1\Delta\rho+\alpha_2|\nabla\rho|^2\right)\mathbf{I}+\alpha_3\nabla\rho\otimes\nabla\rho
+\alpha_4\nabla\nabla\rho\nonumber\\
&+&\alpha_5(\nabla\cdot \mathbf{v})\mathbf{I}+\alpha_6 \mathbf{L},
\label{constequationT}\\
\mathbf{q}&=&q^{(1)}\nabla\varepsilon+q^{(2)}\nabla\rho,
\label{constequationq}
\end{eqnarray}
where $p$, $\alpha_i$ ($i=1,\ldots,6$) and $q^{(i)}$ ($i=1,2$) are suitable material functions depending on the mass density $\rho$ and the internal energy $\varepsilon$.
{As far as the constitutive equation (\ref{constequationT}) is concerned, we observe that the stress tensor $\mathbf{T}$ splits into a standard Navier–Stokes part for compressible fluids 
($\mathbf{T}^{ns}$) and a capillarity stress ($\mathbf T^c$),  characterized
by the following response functions \cite{Heida-Malek}
\begin{equation}
\begin{aligned}
&\mathbf{T}^{ns}=-p\mathbf{I}+\alpha_5(\nabla\cdot \mathbf{v})\mathbf{I}+\alpha_6 \mathbf{L},\\
&\mathbf{T}^c=\left(\alpha_1\Delta\rho+\alpha_2|\nabla\rho|^2\right)\mathbf{I}+\alpha_3\nabla\rho\otimes\nabla\rho+\alpha_4\nabla\nabla\rho,
\end{aligned}
\end{equation}
wherein  $\alpha_i$ ($i=1,\ldots,4$) are material moduli ($[\alpha_1]=[\alpha_4]=Kg\, m^{-1}\, s^{-2}$, $[\alpha_2]=[\alpha_3]=Kg^{-1}m^7 s^{-2}$), $\alpha_5 $ and $\alpha_6$ are viscosity coefficients ($[\alpha_5]=[\alpha_6]=Kg\, m^{-1}s^{-1}$).}\\
 To proceed further, let us expand the specific entropy $s$ around the homogeneous state (where all gradients vanish), retaining only the lower order terms in the gradients entering the state space, \emph{i.e.},
\begin{equation}
\label{specific_entropy}
\begin{aligned}
s &=s_0(\rho,\varepsilon)+s_1(\rho,\varepsilon)|\nabla\rho|^2 +s_2(\rho,\varepsilon)\nabla\rho\cdot \nabla\varepsilon +s_3(\rho,\varepsilon)|\nabla\varepsilon|^2\\
&+s_4(\rho,\varepsilon) \nabla\cdot\mathbf{v}+s_5(\rho,\varepsilon) \Delta\rho,
\end{aligned}
\end{equation}
where $s_i(\rho,\varepsilon)$ ($i=0,\ldots,5$) are some functions of the indicated arguments; $s_0$ represents the equilibrium entropy defined for homogeneous states. Clearly, this expression is not the most general representation of the entropy density as an isotropic scalar function.
 
On the basis of the assumptions on the constitutive relations (\ref{constequationT}), 
(\ref{constequationq}) and (\ref{specific_entropy}) for the stress tensor $\mathbf{T}$ ,  
the heat flux $\mathbf{q}$, and the specific entropy $s$, respectively, the thermodynamical constraints 
we obtain constitute a large set of partial differential equations that we manage using some routines written in the Computer Algebra System Reduce \cite{Reduce}; we are able to solve the
restrictions (\ref{restrhighest}) so finding $s_i(\rho,\varepsilon)\equiv 0$ ($i=2,\ldots,5$), and that $s_1$ is independent of $\varepsilon$, whence the following expression for the specific entropy is recovered:
\begin{equation}
s=s_0(\rho,\varepsilon)+s_1(\rho)|\nabla\rho|^2.
\end{equation}
The principle of maximum entropy at the equilibrium is granted if $s_1(\rho)\le 0$.
It is worth of being remarked that the dependence of the specific entropy $s$ on the modulus of the mass density gradient, here determined from the compatibility with second law of thermodynamics, is in agreement with the constitutive model introduced in Ref.~\cite{morro} where the free energy, with a similar constitutive dependence, has been postulated \emph{a priori}.

From the relations (\ref{restrhigherv})--(\ref{restrhigherrho}) we are able to obtain the entropy flux, 
\begin{equation}
\label{flux_entr}
\mathbf{J}=\mathbf{q}\frac{\partial s_0(\rho,\varepsilon)}{\partial \varepsilon}+2\rho^2 s_1(\rho)(\nabla\cdot\mathbf{v})\nabla\rho.
\end{equation}
Moreover, the following expressions for the material functions entering the Cauchy stress tensor are determined:
\begin{equation}\label{const_eq}
\begin{aligned}
&p(\rho,\varepsilon)=-\rho^2\frac{\partial s_0}{\partial \rho}\left(\frac{\partial s_0}{\partial \varepsilon}\right)^{-1},\\
&\alpha_1(\rho,\varepsilon)=-2\rho^2 s_1\left(\frac{\partial s_0}{\partial \varepsilon}\right)^{-1},\\
&\alpha_2(\rho,\varepsilon)=-\rho\left(\frac{\partial s_0}{\partial \varepsilon}\right)^{-1}\left(\rho\frac{\partial s_1}{\partial \rho}+2s_1\right),\\
&\alpha_3(\rho,\varepsilon)=2\rho s_1\left(\frac{\partial s_0}{\partial \varepsilon}\right)^{-1},\\
&\alpha_4=0.
\end{aligned}
\end{equation}
{
From the thermodynamic restrictions above, it clearly emerges  that the non--local character of  the specific entropy is necessary in order the stress tensor $\mathbf{T}$ to depend on the gradients of the mass density.
Such a result cannot be achieved if one applies the classical Liu procedure; in fact,  in Ref.~\cite{CGOP-2020}, for a non--viscous Korteweg fluid,  it has been shown that without using the extended Liu procedure the specific entropy and the Lagrange multipliers result independent of the gradients entering the state space, and the Cauchy stress tensor cannot depend on the spatial derivatives of the mass density. 
To see that the same result is achieved also for a viscous Korteweg fluid, let us apply the classical Liu method and use only the balance equations (\ref{equations}) for the exploitation of the entropy inequality, \emph{i.e.},
\begin{equation}
\begin{aligned}
&\rho\left(\frac{\partial s}{\partial t}+\mathbf{v}\cdot\nabla s\right)+\nabla\cdot \mathbf{J} -\lambda^{(1)} \mathcal{E}^{(1)}-\boldsymbol\lambda^{(2)}\cdot \mathcal{E}^{(2)}
 -\lambda^{(3)} \mathcal{E}^{(3)}\ge 0,
\end{aligned}
\end{equation}
that, expanded by the chain rule, reads as
\begin{equation}
\label{entropyconstrained:sf}
\begin{aligned}
&\left(\rho\frac{\partial s}{\partial \rho}-\lambda^{(1)}\right)\rho_{,t}-
\rho\lambda^{(2)}_iv_{i,t}+ \rho\left(\frac{\partial s}{\partial \varepsilon}-\lambda^{(3)}\right)
\varepsilon_{,t}\\
&\quad+\rho \frac{\partial s}{\partial \rho_{,k}}\rho_{,kt}+ \rho\frac{\partial s}{\partial v_{i,k}}v_{i,kt}+ \rho\frac{\partial s}{\partial \varepsilon_{,k}}\varepsilon_{,kt} 
+\rho\frac{\partial s}{\partial \rho_{,ik}}\rho_{,ikt}\\
&\quad +\left(\rho v_j\frac{\partial s}{\partial \rho_{,k\ell}}+\frac{\partial J_j}{\partial \rho_{,k\ell}}+\lambda
^{(2)}_i\frac{\partial T_{ij}}{\partial\rho_{,k\ell}}-\lambda^{(3)}\frac{\partial q_j}{\partial\rho_{,k\ell}}\right)
\rho_{,jk\ell}\\
&\quad +\left(\rho v_j\frac{\partial s}{\partial v_{k,\ell}}+\frac{\partial J_j}{\partial v_{k,\ell}}+\lambda
^{(2)}_i\frac{\partial T_{ij}}{\partial v_{k,\ell}}-\lambda^{(3)}\frac{\partial q_j}{\partial v_{k,\ell}}\right)v_{k,j\ell}\\
&\quad +\left(\rho v_j\frac{\partial s}{\partial \varepsilon_{,k}}+\frac{\partial J_j}{\partial \varepsilon_{,k}}+\lambda^{(2)}_i\frac{\partial T_{ij}}{\partial\varepsilon_{,k}}-\lambda^{(3)}\frac{\partial q_j}
{\partial\varepsilon_{,k}}\right)\varepsilon_{,jk}\\
&\quad +\left(\rho v_j\frac{\partial s}{\partial \rho_{,k}}+\frac{\partial J_j}{\partial \rho_{,k}}+\lambda^{(2)}
_i\frac{\partial T_{ij}}{\partial\rho_{,k}}-\lambda^{(3)}\frac{\partial q_j}{\partial\rho_{,k}}\right)\rho_{,jk}\\
&\quad +\left(\rho v_j\frac{\partial s}{\partial \rho}+\frac{\partial J_j}{\partial \rho}-\lambda^{(1)}v_j+\lambda^{(2)}_i\frac{\partial T_{ij}}{\partial\rho}-\lambda^{(3)}\frac{\partial q_j}{\partial\rho}\right)
\rho_{,j}\\
&\quad +\left(\rho v_j\frac{\partial s}{\partial \varepsilon}+\frac{\partial J_j}{\partial \varepsilon}+\lambda
^{(2)}_i\frac{\partial T_{ij}}{\partial\varepsilon}-\rho\lambda^{(3)}v_j-\lambda^{(3)}\frac{\partial q_j}
{\partial\varepsilon}\right)\varepsilon_{,j}\\
&\quad-\left(\rho\lambda^{(1)}\delta_{ij}+\rho\lambda^{(2)}_iv_j-\lambda^{(3)}T_{ij}
\right)v_{i,j} \ge 0.
\end{aligned}
\end{equation}
The latter is a scalar--valued function which is a linear polynomial in the 
derivatives $\rho_{,t}$, $v_{i,t}$, $\varepsilon_{,t}$, $\rho_{,kt}$, $v_{i,kt}$, $\varepsilon_{,kt}$,
$\rho_{,ikt}$, $\rho_{,jk\ell}$, $v_{i,jk}$ and $\varepsilon_{,jk}$ with coefficients depending at most on the field and state variables (\ref{state space}). Since these derivatives are independent of the elements of the state space and can assume arbitrary values, their coefficients must vanish, so providing the following set of thermodynamic restrictions:
\begin{equation}
\label{constraintKorteweg}
\begin{aligned}
&\lambda^{(1)}=\rho\frac{\partial s}{\partial\rho},\qquad  \lambda^{(2)}_i=0,\qquad
\lambda^{(3)}=\frac{\partial s}{\partial \varepsilon},\\
&\frac{\partial s}{\partial\rho_{,k}}=\frac{\partial s}{\partial v_{k,\ell}}=\frac{\partial s}{\partial\varepsilon_{,k}}=\frac{\partial s}{\partial\rho_{,k\ell}}=0,\\
&\left\langle\frac{\partial J_j}{\partial \rho_{,k\ell}}-\frac{\partial s}{\partial \varepsilon}\frac{\partial q_j}
{\partial\rho_{,k\ell}}\right\rangle_{(jk\ell)}=0,\\
&\left\langle\frac{\partial J_j}{\partial v_{k,\ell}}-\frac{\partial s}{\partial \varepsilon}\frac{\partial q_j}
{\partial v_{k,\ell}}\right\rangle_{(jk\ell)}=0,\\
&\left\langle\frac{\partial J_j}{\partial \varepsilon_{,k}}-\frac{\partial s}{\partial \varepsilon}\frac{\partial q_j}
{\partial\varepsilon_{,k}}\right\rangle_{(jk)}=0.
\end{aligned}
\end{equation}
Finally, the residual entropy inequality reduces to
\begin{equation}
\label{entropyresidual}
\begin{aligned}
&\left(\frac{\partial J_j}{\partial \rho}-\frac{\partial s}{\partial\varepsilon}\frac{\partial q_j}
{\partial\rho}\right)\rho_{,j}
-\left(\rho^2\frac{\partial s}{\partial \rho}\delta_{ij}-\frac{\partial s}{\partial \varepsilon}T_{ij}
\right)v_{i,j}\\
&+\left(\frac{\partial J_j}{\partial \varepsilon}-
\frac{\partial s}{\partial\varepsilon}\frac{\partial q_j}{\partial\varepsilon}\right)\varepsilon_{,j}
+\left(\frac{\partial J_j}{\partial \rho_{,k}}-\frac{\partial s}{\partial\varepsilon}\frac{\partial q_j}
{\partial\rho_{,k}}\right)
\rho_{,jk} \ge 0.
\end{aligned}
\end{equation}
Therefore, due to the thermodynamic restrictions (\ref{constraintKorteweg}), the specific entropy and the Lagrange multipliers are independent of the gradients entering the state space; furthermore, by using the constitutive equations 
(\ref{constequationT})--(\ref{constequationq}), and expanding the residual entropy inequality (\ref{entropyresidual}), we obtain a 
polynomial of grade three in the state space variables; necessary conditions for the validity of such an inequality consist in vanishing linear and cubic terms; as a consequence, the Cauchy 
stress tensor turns out to be independent of the spatial derivatives of the mass density. In other words, the classical Liu procedure leads to the conclusion that Korteweg fluids do not exist in nature, since they are not in accordance with second law of thermodynamics. Of course, such a conclusion does not reflect the true physical reality, since Korteweg fluids are relevant in several physical phenomena like capillarity effects. This implies  that such a consequence is an artifact  of the application of the classical Liu procedure. From the technical point of view, we observe that this result is due to the circumstance that 
the terms $\displaystyle\rho \frac{\partial s}{\partial \rho_{,k}}\rho_{,kt}$, $
\displaystyle\rho\frac{\partial s}
{\partial v_{i,k}}v_{i,kt}$, $\displaystyle\rho \frac{\partial s}{\partial \varepsilon_{,k}}\varepsilon_{,kt}
$, and $\displaystyle\rho \frac{\partial s}{\partial \rho_{,ik}}\rho_{,ik t}$ enter the entropy inequality as 
singletons, since the balance equations do not contain similar terms which can be coupled with them. Such a  circumstance leads to the necessary consequence that the quantities $\displaystyle\frac{\partial s}{\partial 
\rho_{,k}}$, $\displaystyle\frac{\partial s}{\partial v_{i,k}}$, $\displaystyle\frac{\partial s}{\partial 
\varepsilon_{,k}}$ and $\displaystyle\frac{\partial s}{\partial \rho_{,ik}}$ must vanish.
The main idea underlying the generalized Liu procedure is to create a generalized Liu inequality containing 
additional terms which can be coupled with the quantities $\displaystyle\rho \frac{\partial s}{\partial 
\rho_{,k}}\rho_{,kt}$, $\displaystyle\rho\frac{\partial s}{\partial v_{i,k}}v_{i,kt}$, $\displaystyle\rho 
\frac{\partial s}{\partial \varepsilon_{,k}}\varepsilon_{,kt}$ and $\displaystyle\rho \frac{\partial s}
{\partial \rho_{,ik}}\rho_{,ik t}$. 
Those terms can only be obtained by introducing into the entropy inequality the spatial gradients of the 
balance equations, up to the order of the gradients entering the state space.  In this way, the entropy density and the Cauchy stress tensor may depend on the spatial gradients of the mass density and, as expected, the Korteweg fluids are fully compatible with second law of thermodynamics.
}

Using the above results, the matrix $\mathbf{A}$ in (\ref{residualquad}) identically vanishes, and the residual entropy inequality becomes
\begin{equation}
\left(\alpha_5 (\nabla\cdot\mathbf{v})^2+\alpha_6 \mathbf{L}\cdot\mathbf{L}\right)\frac{\partial s_0}{\partial \varepsilon}+\mathbf{q}\cdot \nabla\left(\frac{\partial s_0}{\partial \varepsilon}\right)\ge 0;
\end{equation}
the latter is satisfied in all the thermodynamic processes if and only if the following conditions hold:
\begin{equation}
\begin{aligned}
&q^{(1)}\frac{\partial^2 s_0}{\partial \varepsilon^2}\geq0,\qquad q^{(2)}\frac{\partial^2 s_0}{\partial \rho\partial \varepsilon}\geq0,\qquad \alpha_5\frac{\partial s_0}{\partial \varepsilon}\geq0,\qquad \alpha_6\frac{\partial s_0}{\partial \varepsilon}\geq0,
\end{aligned}
\end{equation}
together with
\begin{equation}
\label{relationq1q2}
q^{(1)}\frac{\partial^2 s_0}{\partial\rho \partial\varepsilon}-q^{(2)}\frac{\partial^2 s_0}{\partial \varepsilon^2}=0.
\end{equation}
Some comments about the constitutive relations so characterized are in order. 
In equilibrium situations, in which the gradients of the field variables (except at most the mass density $\rho$) vanish, let us define the absolute temperature $\theta$ by the classical thermodynamical relation $\displaystyle\frac{1}{\theta}=\frac{\partial s_0}{\partial\varepsilon}$. 
{By considering the last relation  as an implicit function, 
\[
G(\rho,\varepsilon, \theta)\equiv\frac{1}{\theta}-\frac{\partial s_0(\rho,\varepsilon)}{\partial\varepsilon}=0, 
\]
under the hypothesis that $\displaystyle\frac{\partial^2 s_0}{\partial^2\varepsilon}\neq 0$, by the implicit function theorem, it follows that  the internal energy $\varepsilon$ can be expressed as a function of  $\rho$ and $\theta$, \emph{i.e.}, $\varepsilon=\varepsilon(\rho,\theta)$.} Thus,  differentiating with respect to $\rho$ the condition
\begin{equation}
\label{condtheta}
\frac{\partial s_0(\rho,\varepsilon(\rho,\theta))}{\partial \varepsilon}-\frac{1}{\theta}=0,
\end{equation}
we get
\begin{equation}
\label{sign}
\frac{\partial^2 s_0}{\partial\rho\partial\varepsilon}+\frac{\partial^2 s_0}{\partial \varepsilon^2}\frac{\partial\varepsilon}{\partial\rho}=0,
\end{equation}
that, used in Eq. (\ref{relationq1q2}), provides
\begin{equation}
\label{relationq1q2bis}
q^{(2)}=-q^{(1)}\frac{\partial\varepsilon}{\partial\rho}.
\end{equation}
Thus, the heat flux reduces to
\begin{equation}
\mathbf{q} = q^{(1)}\frac{\partial\varepsilon}{\partial\theta}\nabla\theta,
\end{equation}
\emph{i.e.}, we have the classical Fourier law of heat conduction. 

Since 
\[
\frac{\partial s_0}{\partial\varepsilon}=\frac{1}{\theta}>0, \qquad\frac{\partial^2 s_0}{\partial\varepsilon^2}=-\frac{1}{\theta^2}\frac{\partial\theta}{\partial\varepsilon}< 0,
\]
it is $\alpha_5\geq 0$ and $\alpha_6\geq 0$, and, as physics prescribes,  $q^{(1)}\leq 0$.
As far as $q^{(2)}$ is concerned, its sign is the same as that of $\displaystyle\frac{\partial\varepsilon}{\partial\rho}$, which in turn, from Eq. (\ref{sign}), has the same sign as  
$\displaystyle\frac{\partial^2 s_0}{\partial\rho\partial\varepsilon}$; thus, due to $\displaystyle\frac{\partial\varepsilon}{\partial\rho}\geq 0$, it is $q^{(2)}\geq 0$. 

Taking into account the definition of absolute temperature,  we can rewrite the entropy flux (\ref{flux_entr}) as
\begin{equation}
\label{entropy_flux_final}
\mathbf{J}=\frac{\mathbf{q}}{\theta}+2\rho^2 s_1(\rho)(\nabla\cdot\mathbf{v})\nabla\rho,
\end{equation}
and we can recognize the contribution of the classical term $\displaystyle \frac{\mathbf{q}}{\theta}$ and an  entropy extra--flux \cite{muller}. We want to stress that the extra--term in the entropy flux arises naturally from the application of the procedure without postulating its existence at the beginning. The entropy extra--flux term in (\ref{entropy_flux_final}) depends on the divergence of velocity ($\nabla\cdot\mathbf{v}$): from a physical point of view, this is a necessary condition, otherwise the Cauchy stress tensor cannot depend on the gradients of mass density, \emph{i.e.}, the essential feature of Korteweg fluids is lost. Let us observe that the generalized procedure here applied, differently from the classical one, is the only method to provide an entropy extra--flux and a Cauchy stress tensor depending on the second order gradients of the mass density \cite{CGOP-2020} without modifying \emph{a priori} the energy balance or the Clausius--Duhem inequality.

In the next Section, analyzing the equilibrium configurations from a mathematical point of view, we obtain a further constraint of the coefficients of the Cauchy stress tensor.

\section{Constraints from equilibrium conditions}
\label{sec:Serrin}

By using the solution to the constitutive functions produced in the previous Section, let us study the form of the phase boundaries at the equilibrium on a purely mechanical framework. 

The search for equilibrium configurations of a Korteweg--like fluid consists in finding solutions of the following constraint:
\begin{equation} 
\label{korteq}
\nabla\left(\left(-p+\alpha_1\Delta\rho+\alpha_2|\nabla\rho|^2\right) \mathbf{I}
+\alpha_3\nabla\rho\otimes\nabla\rho+\alpha_4\nabla\nabla\rho\right)=\mathbf{0},
\end{equation}
where $p$, $\alpha_i$ ($i=1,\dots,4$) are given in (\ref{const_eq}).

Remarkably, in Ref.~\cite{serrin-1983}, Serrin established that, unless rather special conditions are satisfied, the only geometric phase boundaries which are consistent with equation~(\ref{korteq}) are either spherical, cylindrical, or planar.

More in detail, relying on a general theorem proved in Ref.~\cite{Pucci}, Serrin was able to show that the constitutive equation (\ref{constequationT}) for the Cauchy stress tensor must be such that the coefficients therein involved have to satisfy the following condition:
\begin{equation}
\label{serricond}
A \equiv b c+\frac{1}{2}\left(c^2-a\frac{\partial c}{\partial \rho}\right)=0,
\end{equation}
where
\begin{equation}
a=\alpha_1+\alpha_4,\qquad b=\alpha_2+\alpha_3,\qquad c=\frac{\partial\alpha_4}{\partial\rho}-\alpha_3.
\end{equation}

From the mathematical point of view, this constraint can  be understood as a consequence of the fact that the 
equilibrium conditions (\ref{korteq}) have three independent components while liquid--vapour phase equilibria are 
determined by just one physical variable, namely the mass density,  \emph{i.e}, the equilibrium system (\ref{korteq}) is overdetermined. 

In this situation, Pucci  \cite{Pucci} has shown that any solution of this system necessarily has only level surfaces with constant mean and Gaussian curvature \cite{MOV,GO-2019}, 
which are either (pieces of) concentric spheres, or concentric circular cylinders, or parallel planes.

From the physical point of view, this  result can be confirmed by the experimental evidence that several (but not all) phase boundaries  have constant mean curvature.
Moreover, it is worth of observing that Eq.~(\ref{serricond}) is not physically necessary, in the sense that, although rather unusual, without the vanishing of $A$  very few equilibrium configurations are allowed.
However,  the generalized Clausius--Duhem inequality (\ref{entropyconstrained}) does not require $A = 0$; as it is cited in  Ref.~\cite{serrin-1983}, there are significant reasons to accept the restriction $A = 0$ for any physically realistic Korteweg fluid.
 
In our model, in the constraint (\ref{serricond}) we can recognize
\begin{equation}
\begin{aligned}
a&=-2\rho^2s_1\left(\frac{\partial s_0}{\partial \varepsilon}\right)^{-1},\\
b&=-\rho^2\frac{d s_1}{d \rho}\left(\frac{\partial s_0}{\partial \varepsilon}\right)^{-1},\\
c&=-2\rho s_1\left(\frac{\partial s_0}{\partial \varepsilon}\right)^{-1},
\end{aligned}
\end{equation}
leading to
\begin{equation}
\label{serrin_result}
s_0(\rho,\varepsilon)=s_{01}(\rho)+s_{02}(\varepsilon),
\end{equation}
where $s_{01}$ and $s_{02}$ are functions of the indicate arguments.

This relation  means that the specific entropy at equilibrium decomposes in two contributions depending, respectively, on the mass density and the internal energy.
Furthermore, condition (\ref{serrin_result}), from relations (\ref{relationq1q2}) and (\ref{relationq1q2bis}), implies that $\varepsilon=\varepsilon(\theta)$, \emph{i.e.}, the internal energy may depend only upon the absolute temperature, and the heat flux becomes
\begin{equation}
\mathbf{q} = q^{(1)}\frac{d\varepsilon}{d\theta}\nabla\theta.
\end{equation}

The procedure here developed extends to the three--dimensional case the results obtained in Ref.~\cite{GOR-2020} in the one--dimensional case, the only difference being that in the multidimensional case we have some more constraints to be imposed; in fact, in the one--dimensional case no restrictions arise when equilibrium configurations are studied.

\section{Conclusions}
\label{sec:conclusions}
In this paper, we have exploited the entropy principle for a  model of a third grade viscous Korteweg fluid through the application of the extended Liu procedure that requires to use as constraints in the entropy inequality both the field equations and some of their gradient extensions up to the order of derivatives of fields entering the state space.
{As a consequence of the procedure, some new
physical properties of this class of materials have been evidenced. Among them, the most important
one is the presence of the first order gradient of the mass density into the constitutive equation for the thermodynamic potential $s$. Let us observe that this  allows the Cauchy stress tensor to depend on the first and second order gradients of the mass density, so rendering Korteweg fluids compatible with second law of thermodynamics. In particular, if smooth processes are possible, such a property is true even in the absence of the hypothesis of interstitial work flux, \emph{i.e.}, we do not need to modify the classical local balance of energy as proposed in Ref.~\cite{DunnSerrin}.
The interstitial work flux is not mathematically necessary, and its introduction could be decided on the basis of suitable experiments or mathematical arguments relying on kinetic theory.
Nevertheless, in a forthcoming paper we plan to investigate the possibility of including extra--terms both in the balance of energy and in the entropy inequality.}

The exploitation of the thermodynamic compatibility of the constitutive equation
for the Cauchy stress tensor which is sufficiently general  to include
the classical one of the Korteweg fluids, allowed us to recover a complete solution of the system of the
thermodynamic restrictions  in three space dimensions; the cumbersome computations have been done
with the help of the Computer
Algebra System Reduce \cite{Reduce}. The obtained constitutive quantities are also compatible with a constraint arising from mechanical equilibrium configurations where one does not restrict the form of the phase boundaries \cite{serrin-1983}.
Remarkably, the application of the extended Liu procedure does not require neither the modification of the energy balance with the inclusion
of extra--terms (like the interstitial working), nor the \emph{a priori} inclusion of an entropy extra--flux. We were able to  determine 
an explicit expression for the material functions involved in  the Cauchy stress tensor and heat flux by expanding the specific entropy around a homogeneous equilibrium, whereupon the structure of the entropy flux has been algorithmically determined. 

Work is in progress to exploit the entropy principle  for a mixture of viscous Korteweg fluids in the multidimensional case, so extending the one--dimensional results given in Ref.~\cite{CGOP-2020}. 

\section*{Acknoledgments}
Work supported by the ``Gruppo Nazionale per la Fisica Matematica'' (G.N.F.M.) of the Istituto Nazionale di Alta Matematica ``F. Severi'', and by MIFT Department of the University of Messina.

\end{document}